\documentclass{elsart}
\usepackage{amsmath}
\usepackage{amssymb}
\usepackage{bm}
\usepackage[numbers,sort&compress]{natbib}

\newcommand* {\vek}[1]{{\ensuremath{\bm{\mathrm{#1}}}}}
\newcommand* {\ee}{\ensuremath{\mathrm{e}}}

\begin{document}

\begin{frontmatter}

\title{Nanospintronics meets relativistic quantum physics: Ubiquity of
Zitterbewegung effects}

\author[uz]{U.~Z\"ulicke\thanksref{uze}},
\author[rw]{R.~Winkler}
and
\author[jb]{J.~Bolte}

\address[uz]{Institute of Fundamental Sciences and MacDiarmid Institute
for Advanced Materials and Nanotechnology, Massey University,
Private Bag 11~222, Palmerston North, New Zealand}

\address[rw]{Department of Physics, Northern Illinois University,
DeKalb, IL 60115}

\address[jb]{Institut f\"ur Theoretische Physik, Universit\"at Ulm,
Albert-Einstein-Allee~11, D-89069 Ulm, Germany}

\thanks[uze]{
Corresponding author.
E-mail: u.zuelicke@massey.ac.nz}

\begin{abstract}
  We present a unified description of \emph{zitterbewegung}-like
  phenomena for electron and hole systems showing Rashba spin
  splitting as well as for electrons in single-layer and bilayer
  graphene. The former class of systems can be interpreted as
  ``nonrelativistic'' whereas the latter are often called
  ``ultrarelativistic'' so that our unified description indicates an
  interesting connection between these two opposite limits.
\end{abstract}

\end{frontmatter}

\emph{Zitterbewegung} (German for trembling motion) is a highly
oscillatory component in the orbital motion of free Dirac particles
\cite{sch30} which is commonly considered to be a quirk of the very
successful Dirac theory of relativistic electron dynamics. Its
physical significance has been discussed controversially over the
years. Suggestions \cite{sch05a} for a possibly observable
\emph{zitterbewegung}-like dynamics of band electrons in solids have
lifted this discussion onto a new level. The Rashba model
\cite{ras60} studied in Ref.~\cite{sch05a} is equivalent to the
Pauli equation, the latter being the nonrelativistic limit of the
Dirac equation. Recently, graphene has become the subject of
intensive research \cite{gei07}. Its electronic structure bears
remarkable analogies with the ultrarelativistic limit of the Dirac
equation \cite{wal47,div84}. \emph{Zitterbewegung}-like phenomena in
graphene were recently studied in Refs.~\cite{kat06, cse06}. Here we
present a unified description for \emph{zitterbewegung}-like
phenomena in the ``nonrelativistic'' Rashba and the
``ultrarelativistic'' graphene cases based on Refs.~\cite{tha92,
win07} which indicates an interesting connection between these
opposite limiting cases.

We analyze the generalized 2D Rashba Hamiltonian
\begin{equation}
  \label{eq:Hrashba}
  H_R^{(n)}
  = i \beta \left( p_-^n \sigma_+ - p_+^n \sigma_- \right)
  = \beta\left( \begin{array}{cc}
    0 & i\, p_-^n \\ -i \, p_+^n & 0
  \end{array} \right),
\end{equation}
where $\beta$ is a system-dependent prefactor, $p_\pm \equiv p_x
\pm i p_y$, $\sigma_\pm \equiv (\sigma_x \pm i\sigma_y)/2$, and
$\sigma_x, \sigma_y$ denote Pauli spin matrices. The Hamiltonian
$H_R^{(1)}$ is the usual Rashba Hamiltonian \cite{ras60}; for $n=3$
we get the Hamiltonian that describes spin splitting in heavy-hole
systems \cite{win03}. We omit here the spin-independent
kinetic-energy term which, in our context, results in a trivial
correction~\cite{win07}. The Hamiltonian $H_R^{(1)}$ is equivalent
to the well-known Pauli spin-orbit coupling \cite{sak67}. The Pauli
equation represents the nonrelativistic limit to the Dirac equation.

Quasi-free electrons in graphene can be described by the effective
2D Hamiltonians
\begin{equation}
  \label{eq:Hgraphene}
  H_g^{(n)}
  = \beta \left( p_-^n \sigma_+ + p_+^n \sigma_- \right)
  = \beta \left( \begin{array}{cc}
    0 & p_-^n \\ \, p_+^n & 0
  \end{array} \right).
\end{equation}
The Hamiltonian $H_g^{(1)}$ describes electrons in single-layer
graphene \cite{wal47,div84}; and $H_g^{(2)}$ is appropriate for
bilayer graphene \cite{mcc06}. Moreover, the Hamiltonian $H_g^{(1)}$
is equivalent to the ultrarelativistic limit of the 2D Dirac
Hamiltonian.
It is remarkable that all our subsequent results hold for both
classes of Hamiltonians, $H_R^{(n)}$ and $H_g^{(n)}$, which
indicates an interesting connection between the nonrelativistic and
the ultrarelativistic limit of the Dirac equation. From a formal
point of view, this result reflects the fact that $H_R^{(n)}$ and
$H_g^{(n)}$ are related to each other by a simple unitary
transformation. In the following, the Hamiltonian is thus simply
denoted by~$H$.

Following Refs.~\cite{tha92, win07}, we express the Heisenberg
velocity operator for $H$ in the form $\vek{v} (t) = \bar{\vek{v}}
(t) + \tilde{\vek{v}} (t)$, where the mean part is $\bar{\vek{v}}
(t) =
\partial H/\partial\vek{p} - \vek{F}$, and the oscillating part
describing the \emph{zitterbewegung}-like motion reads
$\tilde{\vek{v}} (t) = \vek{F} \, \ee^{-i \hat{\omega} (\vek{p}) \,
t}$. Here the amplitude operator $\vek{F}$ and the frequency
operator $\hat{\omega} (\vek{p})$ are given by
\begin{equation}
  \label{eq:amp_op}
  \vek{F}
  = \frac{\partial H}{\partial \vek{p}} - \frac{n\vek{p} H}{p^2}
  \equiv \sigma_z\, \vek{e}_z \times \vek{p} \, \frac{n H}{i p^2},
  \hspace{3em}
  \hat{\omega} (\vek{p}) = \frac{2 H}{\hbar},
\end{equation}
where $\vek{e}_z$ is a unit vector perpendicular to the 2D plane.
Note that $\vek{F}$ depends explicitly on the index $n$. The
Heisenberg position operator $\vek{r}(t)$ can be expressed in a
similar way as $\vek{r} (t) = \bar{\vek{r}} (t) + \tilde{\vek{r}}
(t)$, where $\bar{\vek{r}} (t) = \vek{r} + \bar{\vek{v}} t + \vek{F}
\:\frac{1}{i \hat{\omega} (\vek{p})}$ and $\tilde{\vek{r}} (t) =
- \vek{F} \: \frac{\ee^{-i \hat{\omega} (\vek{p}) t}} {i
\hat{\omega} (\vek{p})}$.
For the particular case of the Rashba Hamiltonian $H = H_r^{(1)}$,
our general expression for $\vek{r}(t)$ agrees with
Refs.~\cite{sch05a, win07}. For single-layer graphene, $H =
H_g^{(1)}$, it reproduces the results in Ref.~\cite{kat06}, and for
bilayer graphene, $H = H_g^{(2)}$, it agrees with Ref.~\cite{cse06}.

Explicit evaluation shows that $\vek{v} (t)$ oscillates with the
frequency $\omega = 2\beta p^n/\hbar$. The oscillations become
arbitrarily slow for $p \rightarrow 0$. We obtain an estimate for
the amplitude of the oscillations from $\tilde{r}^2 (t) = (n/2)^2 \,
\lambda_B^2$, where $\lambda_B = \hbar/p$ is the de Broglie wave
length. Note that this expression is independent of the prefactor
$\beta$ and it diverges in the limit $p \rightarrow 0$. In a similar
way, we obtain an estimate for the magnitude of the velocity of the
oscillatory motion from
\begin{equation}
  \label{eq:velH2}
  \tilde{v}^2 (t) = \bigl(n \beta p^{n-1} \bigr)^2 =
  \left({\textstyle\frac{n}{2}} \omega \lambda_B \right)^2 .
\end{equation}
The components of $\vek{v}(t)$ do not commute and we have
\begin{subequations}%
\begin{equation}
  \label{eq:cvelH}
  [v_x(t), v_y(t)] =
  2i \bigl(n \beta p^{n-1} \bigr)^2 \,
  \sigma_z \ee^{- i \hat{\omega} (\vek{p}) t},
\end{equation}
which implies the uncertainty relation
\begin{equation}
  \label{eq:uncert}
  \Delta v_x \, \Delta v_y \ge \bigl(n \beta p^{n-1} \bigr)^2 .
\end{equation}
\end{subequations}
Thus both the magnitude of the oscillations of $\vek{v}(t)$ and the
minimum uncertainty for a simultaneous measurement of $v_x$ and
$v_y$ are given by the same combination of parameters \cite{win07}.

\emph{Zitterbewegung}-like phenomena are manifested also by
oscillations of the $z$-components of orbital angular momentum $L_z
(t) = [\vek{r}(t) \times \vek{p}]_z$ and (pseudo)spin $S_z (t)$. We
get
\begin{subequations}%
  \vspace{-3ex}
  \label{eq:angH}%
  \begin{eqnarray}%
   L_z (t) & = & L_z + \frac{n\hbar\, \sigma_z}{2}
    \left(1 - \ee^{-i \hat{\omega} (\vek{p}) t} \right) \, ,
    \label{eq:orbangH} \\
   S_z (t) & = & \frac{n\hbar \, \sigma_z}{2}\: 
    \ee^{-i \hat{\omega} (\vek{p}) t} . \label{eq:spinH}
  \end{eqnarray}
The appearance of the index $n$ in Eq.\ (\ref{eq:spinH}) reflects
the fact that the 2D Hamiltonian $H^{(n)}$ describes effective
particles with pseudospin $z$-component $\pm n/2$. Electrons in the
linear Rashba model as well as electrons in single-layer graphene
have pseudospin $\pm 1/2$, electrons in bilayer graphene are
characterized by a pseudospin $\pm 1$, and heavy holes described by
the cubic Rashba model have pseudospin $\pm 3/2$.
We note also that for $H=H_r^{(1)}$, Eq.\ (\ref{eq:spinH})
represents the well-known spin precession in the effective magnetic
field of the Rashba term, which has been observed
experimentally~\cite{cro05}.
The time dependence in Eqs. (\ref{eq:orbangH}) and (\ref{eq:spinH})
reflects the fact that $L_z$ and $S_z$ are not individually
conserved. However, as expected for models of free quasiparticles,
the total angular momentum $J_z$ is conserved
\begin{equation}
   J_z (t) = J_z = L_z + S_z \; , \label{eq:totangH}
\end{equation}
\end{subequations}
which reflects the fact that $[J_z,H]=0$.

In conclusion, we have presented a unified description of
\emph{zitterbewegung}-like phenomena for electron and hole systems
showing Rashba spin splitting as well as for electrons in
single-layer and bilayer graphene. The former class of systems can
be interpreted as ``nonrelativistic'' whereas the latter are often
called ``ultrarelativistic'' so that our unified description
indicates an interesting connection between these two opposite
limits. 

The authors appreciate stimulating discussions with B.~Trauzettel.
UZ is supported by the Marsden Fund Council from Government funding,
administered by the Royal Society of New Zealand.


\end{document}